\documentclass[12pt]{article} % use larger type; default would be 10pt
\usepackage{graphicx}
\usepackage[title]{appendix}

\begin{document}

\title{A Bayesian Redesign of the \\ First Probability/Statistics Course}
\author{Jim Albert \\Department of Mathematics and Statistics \\ Bowling Green State University}

\maketitle

\begin{abstract}

The traditional calculus-based introduction to statistical inference consists of a semester of probability followed by a semester of frequentist inference.  Cobb (2015) challenges the statistical education community to rethink the undergraduate statistics curriculum.  In particular, he suggests that we should focus on two goals:  making fundamental concepts accessible and minimizing prerequisites to research.  Using five underlying principles of Cobb, we describe a new calculus-based introduction to statistics based on simulation-based Bayesian computation.

\end{abstract}

\section{Introduction}

\subsection{Introductions to Statistics}

At many American universities, a number of introductory statistics courses are taught at the undergraduate level in several departments.  These courses can be divided into three general types.  One type is the introductory statistics course taught to satisfy a quantitative literacy requirement.  This course has a low prerequisite (typically college algebra) and is intended to introduce the student to the science of statistical practice.  This course covers basic tenets of exploring data, sampling distributions, and statistical inference.  A second type of course is the ``methodological" statistics course taught in applied science departments.  This course also has a relatively low prerequisite, but the focus is to introduce the student to the particular statistical methods that may be used in the student's research.  For example, a psychology student might be introduced to statistical inference procedures for means and proportions and regression modeling.  

The third type of course, the focus of this paper, is the two-semester probability and statistics course taught to students in a mathematics or mathematics and statistics department.  In our department, the prerequisite is multivariate calculus.  This course provides an introduction to calculus-based probability and statistical inference.  The text  for the author's probability and statistics course (taken over 40 years ago) was Mendenhall and Schaeffer and the current 7th edition of the text (Wackerly, Mendenhall and Schaeffer (2008)) remains popular.  Table 1 displays the titles of the 16 chapters of the Wackerly et al  (2008) text.  The first half of the course is a traditional introduction to probability including discrete, continuous and multivariate distributions.  The chapters on functions of random variables and sampling distributions naturally lead into statistical inference.  The inferential material includes point and hypothesis testing, regression models, design of experiments and ANOVA models, and categorical and nonparametric methods.  Note that the current edition of the text includes a final chapter introducing Bayesian methods.

\begin{table}[htp]
\caption{Outline of a traditional calculus-based probability and statistics course}
\begin{center}
\begin{tabular}{|c|l|c|l|}
\hline
Chapter & Title & Chapter & Title \\ \hline
1&What is Statistics?& 9& Point Estimation\\ \hline
2&Probability &10&Hypothesis Testing\\ \hline
3&Discrete Random Variables&11&Linear Models\\ \hline
4&Continuous Variables&12&Designing Experiments\\ \hline
5&Multivariate  Distributions&13&Analysis of Variance\\ \hline
6&Functions of Random Variables&14&Categorical Data\\ \hline
7&Sampling Distributions&15&Nonparametric Statistics\\ \hline
8&Estimation&16&Bayesian Methods\\ \hline
\end{tabular}
\end{center}
\label{default}
\end{table}%

\subsection{Concerns About the Traditional Calculus-Based Course}

For some students, this traditional calculus-based probability/inference course may be the first introduction to statistical thinking.  Thinking about the potential audience, there are a number of concerns with the syllabus for this traditional course as laid out in Table 1.

\subsubsection*{Introduction to modern statistics?}

Looking at the chapter titles in Table 1, there seems to be a disconnect of the inferential material with modern statistics.  There is little discussion of methods for exploring data and graphical representation, although these exploration activities are important for the modern statistician.  Although there is some benefit in discussing methods of estimation (such as maximum likelihood and method of moments) and optimal inference (such as the concept of a best hypothesis test), there is little discussion of statistical learning and simulation-based inferential methods popular in modern statistics.  The inferential topics in Wackerly et al (2008) are essentially the same as the topics discussed in the first edition of the text.

\subsubsection*{What does a one-semester student learn?}

Some of the issues described above become more relevant for the student who is able to take only one semester of this probability/inference course.  A semester of probability is certainly fine for many purposes but the student will not be introduced to any data analysis or inferential procedures in this single semester.

\subsubsection*{The preparation for high school teachers}

Many of the students in the calculus-based probability/inference course are prospective math teachers at the secondary education level.   They need to be trained in concepts of probability and statistics that are described in a set of standards for each state.  For example, the state of Ohio has a document ``Ohio's Learning Standards for Mathematics" that states explicitly what skills and knowledge that students should attain at each grade level from K through 12th grade.  Looking at the list of skills, one sees an emphasis on techniques of data analysis in the early grades, concepts of probability are introduced in the middle grades (grades 6 through 8), and  topics in statistical inference are introduced in the final two years of schooling.  It is clear that the prospective math teachers need to have a solid introduction to methods of data analysis with less attention to some of the probability and inferential topics shown in Table 1.   It appears that these math education majors are not well served by the traditional probability and statistics course.

\subsubsection*{A suitable second course?}

Many statistics educators have realized that the traditional calculus-based probability/inference course does not sufficiently cover all of the elements of ``modern" statistics.  So there has been some discussion about an appropriate second course that would follow the probability/inference course. Some possible second courses are regression, categorical data analysis, multivariate statistical analysis, machine learning, nonparametric statistics, design of experiments, and statistical computing/computational statistics.

\subsubsection*{Elements of data science?}

The traditional probability/inference course may use software in the computation of probabilities or the implementation of statistical procedures, but there is typically little attention in this course to statistical programming.  In contrast, due to the availability of ``big data"  and the interest in data science, the modern statistician needs to be fluent in statistical programming.  In a typical exercise, one programs using a language such as R or python to import large datasets,  perform various operations such as define new variables, filtering,  arranging so that the data is in suitable form, and then implementing a suitable statistical analysis.

\subsection{Guiding principles in undergraduate statistics instruction}

The development of any new statistics course should be consistent with  current thinking of the  faculty dedicated to teaching statistics at the undergraduate level.
Recently, the American Statistical Association (ASA) commissioned a workgroup to formulate guidelines for undergraduate instruction in statistics.  The report (Chance et al (2014)) presents the following five principles that should be recognized in the development or revision of undergraduate programs in statistics.

\begin{itemize}
\item The scientific method and the statistical problem solving cycle.

Students should be exposed to all steps of the scientific method in solving statistical problems.  That includes formulating the question, collecting appropriate data to address the question, the exploration stage where one performs the statistical analysis, and the communication of the findings.

\item Real applications

Students should learn statistics through exposure to real data, that is, data that has been collected to learn about an authentic and relevant applied problem.

\item Focus on problem solving

In statistics, it is common to focus on teaching statistical procedures, but the instruction should also be on teaching general principles that will allow students to ask questions and learn new statistical methods in the future.  

\item Increasing importance of data science

Given the new interest in data science, statistical instruction needs to show an awareness of the easily accessibility of large datasets and some of the opportunities in using this data in teaching principles of statistics.

\item Creative approaches to new curricular needs

The ASA believes that there should be opportunity to offer statistical education to a variety of programs in different disciplines.  So there should be flexibility in terms of content, level, and instructional method depending on the needs of the students.

\end{itemize}

\subsection{Cobb's five imperatives}

George Cobb (2015) recently wrote an influential paper that argues that we need to deeply rethink our undergraduate statistics curriculum from the ground up.  Towards this general goal, 
Cobb  proposes the following ``five imperatives" that will hopefully help the process of creating this new curriculum.

\subsubsection*{Imperative 1:  Flatten prerequisites}

Cobb argues that the traditional course in mathematical statistics is the final class of a five-course sequence (Calc 1 to Calc 2 to Calc 3 to Probability to Mathematics Statistics) and this really limits the enrollments in statistics.  In our consulting work, we don't believe it is necessary to take multiple courses in the applied discipline in order to provide statistical help.  Instead, we apply a "just in time`` approach where we learn what we need to know about the applied field to provide the service to the client.  A similar just-in-time approach can be used in teaching a statistics course.  This approach would encourage faculty to think about the necessary skills and concepts and think creatively in the design of new courses.

\subsubsection*{Imperative 2:  Seek depth}

Here Cobb is saying that it is desirable to strip away the technical details so that the student can see the fundamental concepts of the discipline.  
At the undergraduate level, it is desirable to communicate central concepts such as the Central Limit Theorem in words in a simple way.

\subsubsection*{Imperative 3: Embrace computation}

Since computation is such an integral part of the statistics discipline, it should be introduced early in introductory classes.  Given the ready availability of numerical methods, Cobb suggests that computation can be used to motivate statistical concepts.  We should not restrict attention to teaching statistical procedures with closed-form recipes.  By using computation freely, the classroom can resemble applied statistical work which routinely uses computation.

\subsubsection*{Imperative 4: Exploit context}

In applied data analysis, the context provides meaning to an abstract statistical procedure. Cobb describes standard uses of context such as interpretation, motivation, and direction.  In addition, he describes the building up of knowledge from first an illustration, then the abstraction.  Cobbs illustrates the usefulness of teaching how to recognize abstract structure in teaching experimental design.

\subsubsection*{Imperative 5: Teach through research}

Cobb states that our job as statistics instructors (using his words)  ``is not to prepare students to use
data to answer a question that matters; our job is to help them use
data to answer a question that matters."  Teach through research means that the students  should be actively involved in the statistics learning cycle.  Cobbs gives different illustrations of research-based learning at different levels from lower-course undergraduate through first-year graduate.

\subsection{Bayesian Texts}

Given the popularity of Bayesian thinking in applied statistics, a large number of Bayesian texts are currently available.  These books have different purposes and target audiences and one can put many of the texts in the broad classes ``introductory Bayes", ``computational Bayes", ``graduate-level Bayes", and ``applied Bayes".

The purpose of the introductory Bayes texts are to introduce statistical inference to a broad audience assuming only knowledge of college algebra.  
A second type of Bayesian text focuses on Bayesian computational algorithms together with software to implement these algorithms.  Other texts focus on the use of Bayesian software such as BUGS and WinBUGS MCMC software for a  variety of Bayesian models.
Graduate-level Bayesian texts provide a broad perspective on Bayesian modeling including detailed descriptions of inference for single and multiparameter models.  
Other Bayesian texts have a more narrow focus communicating to an audience in a particular applied discipline.  These books don't describe the Bayesian models in mathematical detail, but they are good in making sense of the Bayesian procedures for applied problems.

\subsection{The Task}

Several general comments can be made based on this introduction.  First, there is a strong desire to redesign the traditional probability and statistics course for students with a calculus background.  Second, there is a need for innovation in statistics instruction as documented by the ASA working group.  
Last, Cobb believes that a radical rethinking of statistics instruction is necessary and focuses on five imperatives that should guide this pedagogical innovation.

There are  good reasons to introducing the Bayesian perspective at the undergraduate level.  First, many people believe that the Bayesian approach provides a more intuitive and straightforward introduction than the frequentist approach  to statistical inference.  Second, given the large growth of Bayesian applied work in recent years, it is desirable to introduce the undergraduate student to the some modern Bayesian applications of statistical methodology.
Specifically it is desirable to teach this first  statistics course from a Bayesian perspective.  Documented by the large number of Bayesian texts, there exists plenty of Bayesian instructional material that one can use in the development of this text.   Also there is an increasing amount of Bayesian computational resources as documented by the large number of Bayesian packages in R. (See the Bayesian Task View on CRAN.)

This paper will describe the components of a Bayesian redesign of this calculus-based statistics course in the context of the recommended guidelines in undergraduate statistics instruction.  We revisit Cobb's five imperatives and discuss how each of these imperatives can be addressed in the Bayesian statistics class.  Recently I had the opportunity to develop a Bayesian thinking course for data scientists. This ``Beginning Bayes" course is reviewed  and it is shown how this course addresses some of the recommended guidelines in statistic instruction.  Last, we describe the current status of our ``Probability and Bayesian Modeling" text.

\section{Bayesian Look at Cobbs' Five Imperatives}

\subsection{Introduction}

Here we revisit the five main principles described by Cobb (2015) in the rethinking process of changing our undergraduate statistics curriculum.  In particular, we describe how these principles are implemented in a Bayesian redesign of the first calculus-based statistics class.

\subsection{Flatten prerequisites}

Traditionally,  a student will take a course in Bayesian inference after he or she has taken a probability and inference course taught from a traditional inferential perspective.  So this graduate course in Bayes is implicitly requiring a prerequisite of a traditional inference course where the student gains some familiarity with statistical procedures such as t-test, ANOVA, and multiple regression.  But this knowledge of traditional inference is certainly not necessary before a Bayesian class.  Actually it could be argued that the mix of knowledge of frequentist and Bayesian procedures can be confusing due to the differing interpretation of  inferential procedures using the two paradigms.

\subsubsection*{Using discrete priors}

The student does need a course providing a foundation of probability theory before taking a Bayesian course, but much of the Bayesian paradigm can be communicated with a limited probability background.  For example, much of the Bayesian material in the introductory texts Berry (1996) and Albert and Rossman (2001) is based on discrete probability distributions.  For example, inference about a single proportion is introduced by means of a discrete prior placed on a set of plausible proportion values.  In a similar fashion, inference about a single mean (sampling variance known), is introduced by means of a discrete prior on a set of values of the population mean.

Even inferential comparison methods can be communicated by the use of discrete distributions.  In the comparison of proportions, one places a discrete prior on each proportion, say $p_1$ and $p_2$, and assuming independence, the joint prior is a discrete distribution over a grid of pairs of proportion values $(p_1, p_2)$.  In the case where one has a strong belief that the proportions are equal, one can adjust the prior probabilities along the diagonal values where $p_1 = p_2$.  Inference is achieved by computing the products  (likelihood times prior) for all points and then normalizing the products to obtain the posterior probabilities.  Figure 1 displays a graphical representation of the prior and posterior distributions for two proportions with a uniform prior and 4 out of 12 successes in the first sample and 6 out of 12 successes in the second sample.

\begin{figure}[htbp]
\begin{center}
\includegraphics[scale = 0.5]{"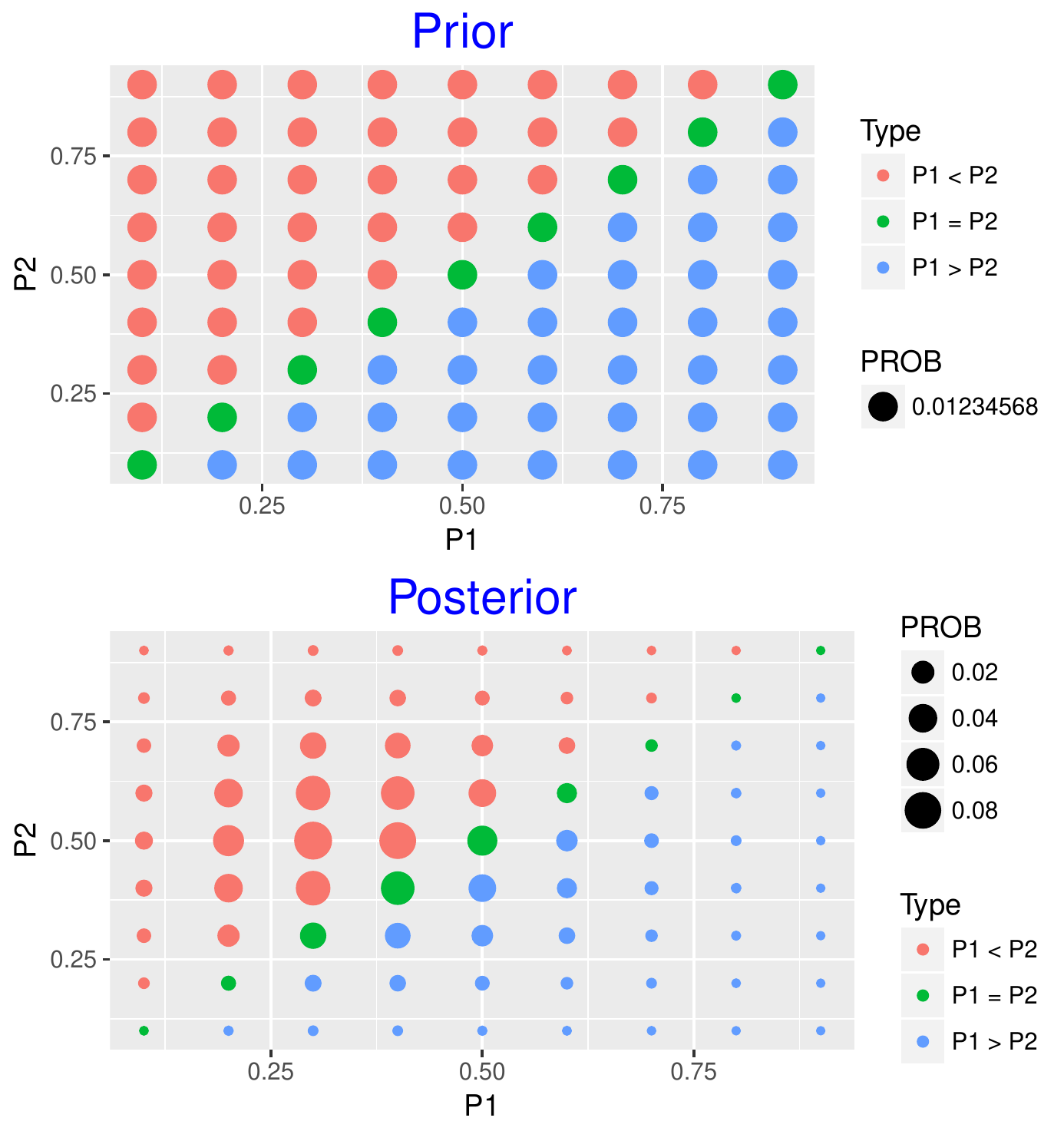}
\caption{Graphs of prior and posterior of two proportions using a discrete prior. }
\end{center}
\end{figure}

\subsubsection*{Using simulation}

Simulation provides another attractive ``flattened prerequisites" strategy in presenting Bayesian inference.  One aspect of Bayesian inference is  that summaries of a posterior distribution are expressible as integrals which may be difficult to evaluate, especially for multiparameter problems.  One way to avoid the integration issue is to simulate a large number of values from the posterior distribution.  Then posterior summaries are expressed as data summaries of the simulated posterior sample.  Given that students typically have some skills in exploring and summarize data, these same skills can be applied to the posterior ``data".

Simulation can be introduced for problems where a conjugate prior of a familiar functional form is available.  For example, using a beta prior with binomial data leads to a beta posterior and summaries of the posterior can be found by taking a simulated sample from a beta distribution.   Other problems may have not have convenient conjugate priors, but Gibbs sampling can provide a straightforward way of simulating a posterior sample by successively simulating from conditional posterior distributions.    For example, in the familiar situation where one is learning about the mean $\mu$ and variance $\sigma^2$ from normal sampling, then assuming a prior proportional to $1 / \sigma^2$ the posterior density is proportional to
$$
\frac{1}{(\sigma^2)^{n/2 + 1}} \exp\left(- \frac{1}{2 \sigma^2} \sum (y_i - \mu)^2\right).
$$
In this setting, by recognizing that the distributions [$\mu | \sigma^2$] and [$\sigma^2 | \mu$] have common functional forms, one can construct a Gibbs sampler based on normal and inverse-gamma sampling.

The Gibbs sampler can be viewed as an introduction to the class of Markov chain Monte Carlo (MCMC) algorithms routinely used by applied scientists in Bayesian analyses.  Once the student gets some basic understanding of Markov chains, then it is desirable to introduce a general class of algorithms such as the Metropolis-Hastings random walk that can be used for a wide variety of inferential problems.

\subsection{Seek depth}

Some introductory statistics texts introduce Bayesian inference, but much of the discussion is devoted to a definition of the prior, likelihood, and posterior and the mechanics of computing the posterior and  inference for some standard examples.  In the calculus-based introductory class, it is desirable to teach  other components of the Bayesian paradigm.

\subsubsection*{Prior elicitation}

One advantage of the Bayes perspective is the opportunity to input expert opinion by the prior distribution.  So it is desirable for the course to include some strategies for constructing one's prior when one has substantive prior information.  In the case where one has little prior knowledge, some discussion about suitable noninformative or vague priors is needed.

\subsubsection*{Sensitivity of inference with respect to model assumptions}

A Bayesian model consists of a choice of prior and sampling density.  The prior is an approximation to a person's ``true" prior that would be obtained after a long period of elicitation, and likewise a particular sampling density is chosen for convenience.  When there is doubt about the accuracy of these approximations, the Bayesian paradigm provides a convenient mechanism to explore the sensitivity of a particular inference to these choices of prior and sampling density.  This course should demonstrate this type of Bayesian sensitivity analysis for different inferential problems.

\subsubsection*{Model checking and model comparison}

Any Bayesian inference is conditional on the model assumptions including prior and sampling density.  
When there are several Bayesian models under consideration, one can use the marginal likelihoods of the models to compare how they fit the data.  Bayes factor, the ratio of the marginal likelihoods, can be used to compare several models.  For a single model, diagnostics can be performed by exploring the posterior predictive density of a testing function.  In an introduction to Bayesian thinking, one should illustrate the use of posterior predictive checking, and also demonstrate the formal selection of models by use of Bayes factors.

\subsubsection*{Hierarchical modeling}

One of the Bayesian success stories is the use of hierarchical or multilevel modeling when one wishes to simultaneously estimates parameters from several groups.  In regression situation, it is common to want to combine regression estimates from several groups, and hierarchical modeling is an effective way to achieve partial pooling of the separate regression estimates.  By use of simple examples such as the simultaneous estimation of several proportions, one can demonstrate the value of these multilevel models.

\subsection{Embrace computation}

In the traditional Bayesian course, much of the material is devoted to the derivation of posterior and predictive distributions for conjugate problems.  Although these derivations are helpful in understanding how the prior information and data are combined in a posterior, they don't introduce the student to the modern MCMC simulation algorithms currently used in Bayesian research.  It is important that the students gain some basic understanding of simulation-based Bayesian computation.

Simulation of posterior and predictive distributions can be introduced in Bayesian computations from the simplest to the most sophisticated models.   There are several advantages to the use of simulation in this context.  Summarizing a sample of simulated draws from the posterior is analogous to the task of summarizing a large sample of data, and so data analysis skills can be applied to the Bayesian inference setting.  It is straightforward to perform inference about a transformation of the parameter vector, say $h(\theta)$.  If one has the sequence of simulated draws  \{$\theta^{(j)}$\}  from the posterior $g(\theta | {\rm data})$, the sample of simulated draws \{$g(\theta^{(j)})$\} will be distributed from the posterior of  $h(\theta)$.  Last, simulation can be used to simulate draws from the posterior predictive distribution.  In a Bayesian model where $y$ is distributed from the sampling model $f(y | \theta)$ and $\theta$ has a prior $g(\theta)$, then a simulated draw from the predictive density is obtained by first simulating $\theta$ from $g$, call the simulated draw $\theta^*$, and then simulating $y$ from the distribution $f(y | \theta^*)$.  One use of simulating replicated samples from the posterior predictive distribution is in model checking.

\subsubsection*{Discrete Bayes}

To illustrate simulation in a basic setting, suppose one is interested in learning about a proportion $p$ and a discrete prior assigns probabilities \{$g(p_j)$\} on the set \{$p_j$\}.  If one observes $y$ successes in $n$ trials, then the posterior probabilities are proportional to the products $ \{g(p_j) \times p_j^y (1 - p_j)^{n - y}\}$.  One simulates from the posterior by sampling with replacement \{$p_j$\} where the sampling probabilities are proportion to $ \{g(p_j) \times p_j^y (1 - p_j)^{n - y}\}$.  On R this can be done with a single application of the \texttt{sample} function.

\subsubsection*{Conjugate priors}

As mentioned earlier, in introducing Bayesian thinking, it is helpful to describe several conjugate problems where the prior and posterior have familiar distributions such as the beta, gamma, or normal. For example, if a beta$(a, b)$ prior is assigned to a proportion $p$ and if $y$ successes and $n - y$ failures are observed, the posterior for $p$ is beta with updated shape parameters $a^* = a + y$ and $b^* = b + n - y$.  A 90\% probability interval can be found by extracting quantiles from a beta($a^*, b^*)$ density.  Alternatively, one can obtain this interval by simulating, say a 1000, draws from a beta($a^*, b^*)$ distribution and finding sample quantiles of the simulated draws \{$p^{(j)}$\}.

One criticism of this simulation approach is that one is introducing error by simulation, and this simulation-based probability interval is only an approximation to the exact probability interval found using  beta quantiles.  However, this simulation approach has several advantages in this conjugate setting.  Simulation is seen as a general strategy for computing posterior distributions, and one can assess the accuracy of simulation-based posterior summaries by comparing the simulation summaries with the exact summaries.  Moreover, one can illustrate the ease of computation of the posterior of a function of the parameter, say the odds  $h(p) = p / (1 - p)$ by simply transforming the simulated values of $p$ from the posterior.

\subsubsection*{Normal approximations}

Once the student has some familiarity with the normal distribution, then by use of the Laplace approximation (Kass and Raftery, 1995), the normal distribution can be seen as a quick and convenient approximation to a posterior distribution of several parameters.  Once this approximation is developed, then one can perform posterior calculations by drawing a large sample from the approximate normal distribution.  

\subsubsection*{Markov chain Monte Carlo methods}

Once the student has been exposed to simulation as a general Bayesian computational tool, then MCMC can be introduced as a general method for simulating a Markov chain in situations where the posterior distribution is in a less-tractable form. 
A Markov chain can be introduced in the simple  setting where one has a finite collection of states.  In continuous-parameter settings, the 
random walk Metropolis algorithm is an attractive method for setting up a Markov chain.

\subsubsection*{Software}

A number of general-purpose software programs are available for Bayesian MCMC computation such as openBUGS, JAGS, Nimble, and Stan and the student should be introduced to the use of one of these programs.  The main task in the use of these programs is the specification of a script defining the Bayesian model and then the Bayesian fitting is implemented by a single function that inputs the model description, the data, and any tuning parameters of the algorithm.

There are several benefits to the use of this general-purpose software.  First, by writing the script defining the full Bayesian model, the student gets a deeper understanding of the sampling and prior components of the model.  Second, by use of this software for sophisticated models such as hierarchical models, this software lowers the bar for students to implement these methods.  The focus of the students' work is not the computation but rather the summarization and interpretation of the MCMC output.

\subsection{Exploit context}

All of the aspects of a Bayesian analysis are communicated best through interesting case studies or extended examples.  In a good case study, one describes the background of the study and the inferential or predictive problems of interest.  A Bayesian class should include several engaging applied examples where one describes the construction of the prior to represent expert opinion, the development of the likelihood (which preferably does not fit a standard form), and the use of the posterior distribution to address the questions of interest.  Indeed one aspect of memorable Bayesian texts are the inclusion of particular examples that help in motivating the description of the methodology.

When I think about notable Bayesian texts, I think about some of the motivating examples.  In particular, Link and Barker (2009) use a number of interesting ecological case studies and applications.  When I reflect on Gelman and Hill (2007), I think about the interesting use of political examples to illustrate multilevel modeling.  McElreath (2015) also has engaging examples -- I use the coffee shop waiting time in my own workshops to illustrate multilevel modeling.  Efron and Morris' example of predicting the end of the season batting averages for 18 players in Efron and Morris (1975) is still used in modern texts to illustrate hierarchical modeling.  Although the data  is over 40 years old, the popularity of this example illustrates the power of a good example in communicating statistical concepts.

\subsection{Teach through research}

Bayesian thinking is best learned in the concept of one's own statistical study.  In a student's own project, he or she will think carefully about the choice of prior that approximately represents his or her beliefs and a likelihood function will be chosen for the particular type of data that is collected.  The student collects data.  By exploration of the posterior, he or she will address the question of interest and compare the conclusion with the prior opinion.

A Bayesian project can be implemented in a statistics class at any level.  Albert (2000) illustrates the use of a sample survey project in an introductory statistics class.  In this project, the student performs a survey to learn about a particular population proportion.  The student uses a discrete prior on the proportion to represent his or her opinion and after collecting data, computes the posterior on the discrete set of proportion values.  This exercise demonstrates the process of adjusting one's opinion after observing data.

Chapter 17 of Gelman and Nolan (2017) describes several activities where the students are engaged in Bayesian thinking.  In Section 17.1, students are asked to construct a 50 percent subjective probability interval for the number of quarters in a jar.  In Section 17.2, students are asked to estimate the kidney cancer death rates for a number of counties in a particular state.  The challenge is to learn about the underlying true cancer rates when one has observed rates from counties based on different sample sizes.  Although these activities are described in a classroom setting, they could be modified to create mini-projects for the students.

\section{{\it Beginning Bayes} Course}

\subsection*{Goals of Course}

DataCamp is a company that offers short courses in topics in data science and modeling using the R and python script languages.  In a typical course requiring approximately four hours of effort, the student will view a series of instructional videos and get practice in applying the concepts using an online system using either R or python.  The students who enroll in DataCamp courses come from a wide range of backgrounds.  Many are employed in positions that require some data science experience and they are taking specific courses to learn particular skills in data science or modeling. The general goal of the ``Beginning Bayes" course was to introduce Bayesian thinking to data scientists assuming only a minimal background in probability.  There was some discussion about DataCamp offering a more sophisticated Bayesian regression course using Stan software and so this course was viewed as a possible introductory Bayesian course that would be a prerequisite to the Bayesian regression course. Using Cobb's terminology, this course is the ultimate ``flattened prerequisites" course requiring little mathematics preparation from the student.

As in all courses offered by DataCamp, the instructional module consists of a series of chapters, where each chapter contains a series of instructional videos and online exercises using the R statistical system.  A R package {\texttt TeachBayes} was written to accompany this course.  The package includes functions for visualizing probability calculations for beta and normal distributions, and functions to facilitate Bayes' rule calculations (such as {\texttt bayesian\_crank} and {\texttt two\_p\_update}).

\subsection*{Course Topics}

\subsubsection*{Topic 1: Introduction to Bayesian thinking}

In this first chapter, discrete probability distributions are introduced using a random spinner defined by a set of spinner areas.  When there are several plausible spinners that are spun, Bayes' rule is used to learn about the spinner identity from the observed spinner values.  In this chapter, the student gets introduced to the notions of prior, likelihood, and posterior.  This chapter concludes with an illustration of sequential learning where the posterior distribution on the spinners after one spin become the prior distribution before another spin.

\subsubsection*{Topic 2:  Learning about a binomial probability}

The second chapter presents the familiar problem of learning about a binomial probability.  It starts with the use of a discrete prior using Bayes' rule.  Then a beta prior is used to represent prior knowledge about a continuous-value proportion and a special R function is used to find the shape parameters of the beta prior that match the knowledge of two percentiles.  The TeachBayes package is used to find probabilities and percentiles for the beta distribution, and Bayesian interval estimates and  tests are found by these summaries. 
Simulation is also introduced as an alternative way of performing posterior calculations.

\subsubsection*{Topic 3: Learning about a normal mean}

The third chapter describes Bayesian inference for a normal mean when the sampling variance is assumed known.  This chapter follows the same basic structure as Chapter 2.  One starts with the use of a discrete prior where one specifies a list of plausible values of the mean, and one computes posterior probabilities by Bayes' rule.  Then the normal prior is introduced, and a R function {\texttt normal\_update} is used to find the posterior mean and standard deviation of the normal posterior.  Simulation is introduced to facilitate posterior computations about the mean, and it is also used to simulate future observations from the predictive distribution.

\subsubsection*{Topic 4: Bayesian comparisons}

The final chapter introduces comparison of proportions, inference about a normal mean when the sampling standard deviation is unknown, and Bayesian regression. For comparing two proportions, a discrete prior over a grid is used to represent prior opinion and inference is another application of Bayes' rule.  Next, independent beta distributions are used to represent opinion when the proportions are continuous-valued.  Simulation is used to learn about the difference in proportions.  The \texttt{arm} package is used to illustrate Bayesian regression calculations.  The normal mean inference problem is a special case of a regression model with only a constant term, and the two-group model is a regression model with an indicator value for the group covariate.  In each case one simulates from a Bayesian regression model with a noninformative prior.  By use of transformations on the matrix of simulated draws, one can perform inference for a normal percentile and a standardized group effect.

\section{Text:  {\it Probability and Bayesian Modeling}}

Monika Hu and I recently completed  (Albert and Hu (2019)) an undergraduate text that is a Bayesian redesign of the calculus-based probability and statistics sequence.  Table 2 gives the table of contents of our text.  I had previously written (Albert (2015)) a data analysis and probability text for prospective middle-school and high school teachers of mathematics and some of the  material  from that text is used for the probability content of our book.

\begin{table}[htp]
\caption{Table of Contents of Proposed Bayesian Course}
\begin{center}
\begin{tabular}{|c|l|l|} \hline
Chapter & Title & Contents \\ \hline
1 & Probability:  A Measurement  & interpretations of probability, \\ 
& of Uncertainty & probability axioms, assigning  \\ 
& & probabilities \\ \hline
2 & Counting Methods &  \\ \hline
3 & Conditional Probability & includes Bayes' rule \\  \hline
 4 & Probability Distributions & focus on coin-tossing \\
  &  & models \\ \hline
  5 & Continuous Distributions & includes normal distribution \\
  & & and Central Limit Theorem \\ \hline
 6 & Joint Probability & discrete and continuous \\
 & Distributions & distributions \\ \hline
 7 & Learning About a & Bayes' with discrete models \\
 & Proportion & beta prior, posterior inference \\ \hline
 8& Learning About a Mean & Bayes' with discrete models \\ 
  & & normal prior, inference and prediction \\ \hline
9 &  Simulation by MCMC & Gibbs sampler, Metropolis-Hastings \\
& & JAGS software \\ \hline
 10 &  Bayesian Hierarchical Modeling &  exchangeable models for \\ 
 &  & means and proportions \\ \hline
 11 &Simple Linear Regression& \\ \hline
12 & Multiple Regression& \\ 
& and Logistic Models& \\ \hline
13&Case Studies& text analysis, hierarchical \\
&  & regression, latent class models  \\ \hline
\end{tabular}
\end{center}
\label{default}
\end{table}%

As can be seen from Table 1, the probability material in Chapters 1 through 6 resembles the material for a traditional probability course including foundations, conditional probability, discrete and continuous distributions. There are some notable omissions of content from the Wackerly et al (2008) probability content.  There is relatively little discussion of distributions of functions of random variables and there is little discussion of the common families of distributions with the exception of the coin-tossing distributions (binomial and negative binomial) and the normal.  For obvious reasons, there is limited discussion of sampling distributions although the Central Limit Theorem is introduced as one application of the normal distribution.

Although there are applications of Bayes' rule in the probability chapters, the main Bayesian inferential material begins in Chapters 7 and 8 with a discussion of inferential and prediction methods for a single binomial proportion and a single normal mean.  The foundational elements of Bayesian are described in these two chapters including the construction of a subjective prior, the computation of the likelihood and posterior distributions, and the summarization of the posterior for different types of inference.  Predictive distributions are described both for predicting future data but also for implementing model checking

Since the remaining material in the text is heavily dependent on simulation algorithms, Chapter 9 provides an overview of Markov Chain Monte Carlo (MCMC) algorithms with a focus on Gibbs sampling and Metropolis-Hastings algorithms.  This chapter begins with an informal discussion of the Metropolis-Hastings random walk to provide some intuition into the logic underlying the algorithm.  Once the random walk algorithm is described, one can discuss all of the implementation issues such as the choices of starting value and number of iterations and MCMC diagnostic methods.  Once there is reasonable confidence that the MCMC sample is an approximation representation of the posterior distribution, then the text describes the use of the MCMC sample for different types of inference. 

The remaining chapters (Chapters 10 through 13) use JAGS as the computational software for illustrating Bayesian thinking for some popular models.  Hierarchical modeling is introduced in Chapter 10 with a focus on simultaneously estimating a set of normal means and a set of binomial proportions.  Bayesian regression models are described in Chapters 11 and 12.  Chapter 11 focus on priors and posterior and predictive computations for the simple linear regression model with a single input.  Chapter 12 extends this model to describe Bayesian multiple regression and logistic regression when the response variable is binary. Chapter 13 describes several case studies including text mining, simultaneously estimating a series of career trajectories using a hierarchical model, and a latent class analysis.

\begin{appendices}
\section{Selection of Bayesian Texts}

\subsubsection*{Introductory Bayes}

The purpose of the introductory Bayes texts are to introduce statistical inference to a broad audience assuming only knowledge of college algebra.  Blackwell (1969) and Schmidt (1969) are a couple of early examples of these introductory Bayes texts.  Much of the material in these early texts assumes that the parameter of interest takes only a discrete collection of values.  Berry (1996) and Albert and Rossman (2001) follow a similar strategy in using discrete models to introduce Bayesian thinking.  

\subsubsection*{Computational Bayes}

A second type of Bayesian text focuses on Bayesian computational algorithms together with software to implement these algorithms.  Albert (2009) and Marin and Robert (2013) focus on computational strategies with illustrations of these Bayesian computations using the R language.  Other texts focus on the use of Bayesian software; for example, Lunn et al (2012) and Ntzoufras (2009) illustrate the use of BUGS and WinBUGS MCMC software for a wide variety of Bayesian models.

\subsubsection*{Graduate-Level Bayes}

Graduate-level Bayesian texts provide a broad perspective on Bayesian modeling including detailed descriptions of inference for single and multiparameter models.  Gelman et al (2013), Carlin and Louis (2008), and Hoff (2009) are good examples of modern texts typically used at the graduate level.  Gill (2014) and Jackman (2009) provide a good foundation of Bayesian theory at the graduate level with applications to the social sciences. 

\subsubsection*{Applied Bayes}

Other Bayesian texts have a more narrow focus communicating to an audience in a particular applied discipline.  For example, Link and Barker (2009) write to an audience in applied ecology research and McElreath (2015) is aimed at researchers in the natural and social sciences.  These books don't describe the Bayesian models in mathematical detail, but they are good in making sense of the Bayesian procedures for applied problems.

\section{Bayesian R Packages (A Personal List)}

\begin{itemize}
\item \texttt {LearnBayes}

\item \texttt {TeachBayes}

\item \texttt {arm}

\item \texttt {rjags}

\item \texttt{rstan}

\item \texttt {rstanarm}

\item \texttt {coda}

\item \texttt {bayesplot}

\end{itemize}

\end{appendices}

\end{document}